%Paper: hep-th/9208053
%From: Ivan Kostov <kostov@poseidon.saclay.cea.fr>
%Date: 21 Aug 92 21:30:49+0200

%%%%%%%%%%%%%%%%%%%%%%%%%%%%%%%%%%%%%%%%%%%%%%%%%%%%%%%%%%%%%%%%%%%%%%%%%%%
\input harvmac
\def\CN{{\cal N}}
\def\mm{\sqrt{\mu}}
\def\m{\mu}
\def\e{\epsilon}
\def\CZ{{\cal Z}}
\def\CS{{\cal S}}

\def\l{\ell}
\def\CA{{\cal A}}

\def\CV{{\cal V}}

\def\e{\epsilon}

\def\p{\partial}

\def\CD{{\cal D}}
\def\R{\relax{\rm I\kern-.18em R}}
\font\cmss=cmss10 \font\cmsss=cmss10 at 7pt
\def\Z{\relax\ifmmode\mathchoice
{\hbox{\cmss Z\kern-.4em Z}}{\hbox{\cmss Z\kern-.4em Z}}
{\lower.9pt\hbox{\cmsss Z\kern-.4em Z}}
{\lower1.2pt\hbox{\cmsss Z\kern-.4em Z}}\else{\cmss Z\kern-.4em Z}\fi}
\def\pl{{\it  Phys. Lett.}}
\def\np{{\it Nucl. Phys. B}}

\def\s{\sigma}
\def\r{{\rm Re}}
\def\i{{\rm Im}}

\def\vp{\varphi}

%\draftmode
\Title{}{\vbox{\centerline{
% Strings with Discrete Target Space}
Gauge Invariant Matrix Model}
\vskip2pt\centerline{
 for the \^A-\^D-\^E Closed Strings }
%as an Eigenvalue Problem }
}}
%   \footnote{}{*optional footnote on title}
%\Title{ }{Strings with Discrete Target Space as an Eigenvalue Problem}
\bigskip\centerline{ I. K. Kostov \footnote{$^\ast $}
{ on leave from the Institute for Nuclear Research and Nuclear Energy,
72 Boulevard Tsarigradsko Shosse, 1784 Sofia, Bulgaria}}
\bigskip\centerline{{\it Service de
 Physique Th\'eorique}  \footnote{$^{\dagger}$}{
Laboratoire de la Direction
 des Sciences de la Mati\`ere du
Commissariat \`a l'Energie Atomique}{\it  de Saclay} }
\centerline{{\it CE-Saclay, F-91191 Gif-sur-Yvette, France}}

%if too many authors for abstract on same page, say   \vfill\eject\pageno0

\vskip .3in

\baselineskip8pt{
 The models of triangulated random surfaces embedded in
(extended) Dynkin diagrams are formulated as a gauge-invariant matrix
model of Weingarten type. The double scaling limit of this model
is described by a collective field theory with nonpolynomial interaction.
 The propagator in this field theory is essentially  two-loop correlator
 in the corresponding string theory.
}
\bigskip
\bigskip
\bigskip
\leftline{Submitted for publication to: {\it Physics Letters B}}
\rightline{SPhT/92-096}
%\draft
\Date{8/92} %replace this line by \draft  for preliminary versions
             %or specify \draftmode at some point

%if you want double-space, use e.g. \baselineskip=20pt plus 2pt minus 2pt

\newsec{Introduction}

The string theories with discrete target space are interesting mainly
because of their interpretation as theories of two-dimensional quantum
gravity with matter fields. The (discrete) degrees of freedom of
 the matter fields are labeled
by the points of the target space.

It appeared that the theories with nontrivial critical behaviour are
classified by the two-dimensional integrable statistical systems that
can be defined on a generic planar graph. The continuum limit of these
models is described by a conformal field theories with central charge
$C \le 1$. The explicit construction involving integrable models
on planar graphs \ref\Iade{I. Kostov, \np 326 (1989) 583} allowed to apply
the formalism called ``Coulomb gas picture'' and map the string theories
with discrete target space onto a loop gas model on a random surface.
The latter can be solved order by order in the topology of the random
surface by cutting it along the loops into elementary pieces
\ref\Idis{I. Kostov, \np 376 (1992) 539}.

Although each step in this construction is quite trivial, as a whole
it seems heavy to follow, because its constant appeal for geometric
 imagination.
% being too ``geometric''.
 This is why we propose
in this letter an alternative, purely algebraic method of solution
using a special large $N$ matrix model. In this way  the
integrable  statistical systems
on triangulated random surfaces will be formulated
in terms of a collective field theory describing the dynamics of the
eigenvalues of the matrix field.

Our motivation is not only the simplicity of presentation.
 The analysis of the SOS
string (target space $\Z$) showed that
 its field content is different than the one  of the string
with continuous target space $\R$. In particular, there are no
``special states'' in the SOS string. This is quite a puzzle because
the discrete chain of matrices is known to be
equivalent to matrix quantum mechanics
after adjusting the parameters \ref\klb{I. Klebanov, String theory in two
dimensions, Lectures at the ICTP Spring School on String Theory
and Quantum Gravity, Trieste, April 1991; Princeton preprint PUPT-1271}

  Our conjecture \Idis\ was that the SOS string
corresponds to a matrix chain with the critical distance between two
subsequent points of the discretized line where a Kosterlitz-Thouless
transition occurs. In this case  the momenta of the
 special states are multiples
of the period in  the momentum space.
The equivalence of the SOS string to a matrix field theory allows to
prove the above conjecture.
%The first models of discretized random surfaces immitated the lattice
%gauge theories. The Weingarten model in $D$ dimensions is obtained from the
%corresponding gauge theory by replacing the $U(N)$ Haar measure with the
%gaussian measure. As a result one obtains again a gauge-invariant matrix
%field theory but with additional radial degrees of freedom which change
%the critical behaviour of the model. Recently, the Weingarten model was
%reconsidered and adapted to dimensions $D \ge 1$.
%We think that the gauge invariance of the Weingart
%en-like models makes them
%more attractive than other matrix models studied in recent times.
%The basic feature of these models is that the matrix field variables
%are associated with the $links$  of the discretized world sheet.
%
\newsec{ The triangulated random surface as a large $N$ matrix field theory}

Our matrix model will be a variant of the Weingarten model \ref\Wg{D.
Weingarten, \pl 90 (1980) 280} which has been recently reconsidered in the
interesting paper of Dalley \ref\Dal{S.Dalley, {\it Mod. Phys. Lett.}
A7 (1992) 1651}.
It describes triangulated random surfaces embedded in a discrete target space
$X$.
 For the moment it is sufficient to assume
that $X$ has a structure of a one-dimensional simplicial complex. In other
words, $X$ is an ensemble of points $x$ and links $l$.
 In particular,
this can be the $D$-dimensional hypercubic lattice considered
 as a collection of sites and
links. However, a nontrivial continuum limit exists only if $D=1$.
More generally,  $X$ can be an
(extended)
Dynkin diagram of A-D-E type \Iade .

 A link having as
extremities the points $x$ and $x'$ will be  denoted by $<xx'>$.
Below we assume that two points are connected by at most one link only to
simplify the notations.
The target space $X$ is defined completely by its connectivity matrix
\eqn\conm{C_{xx'}=({\rm the \ number \ of \ links \ connecting \ }
x \ {\rm and} \  x')}
We are going to consider only embeddings compatible with the structure
of one-dimensional simplicial complex. This means that the triangulated
surface $\CS$ is embedded so that the images $x(\s)$ and $x(\s ')$ of each
pair of
nearest neighbour points $\s , \s ' \in \CS$ either coincide, or are
connected by a link in $X$. Under the assumption that
 that the space $X$ does not contain
loops of length 3, each  triangle $\bigtriangleup _{\s \s ' \s ''}$
of $\CS$ is mapped  onto a single point $x$ or a link $<xx'>$ of $X$.

The entities of the  matrix model will be $N \times N$ matrices associated
with the points $and$ links of the target space $X$.
We introduce a $complex$ matrix field variable
$A^{<xx'>}_{jk} = (A^{<x'x>}_{kj}$ for each  link $<xx'>$ of the target space,
and a Hermitean matrix variable $\Phi_{x}$ for  each point $x \in X$.
The partition function is defined as
\eqn\partf{\CZ=\int \prod _{<xx'>} dA^{<xx'>}dA^{<x'x>} \prod_{x}d\Phi^{(x)}
e^{-\CA [A,\Phi]}}
\eqn\actn{\eqalign{
\CA [A,\Phi]=& N \tr \big[ {1 \over 2} \sum _{<xx'>} A^{<xx'>}A^{<x'x>}
+ {1 \over 2} \sum _{x}[\Phi^{(x)}]^{2}
 -  {1 \over 3 } \kappa _{0} \sum _{x} [ \Phi^{(x)}]^{3}\cr
&+{1\over 2} \kappa_{1} \sum_{<x,x'>}\big( A^{<xx'>}A^{<x'x>} \Phi ^{(x)}
+ A^{<xx'>} \Phi ^{(x')}A^{<x'x>} \big)  \big] \cr}}
The free energy  $\CF=\log \CZ$ of the model is expressed as a sum over all
connected triangulated surfaces $\CS$ embedded in $X$:
\eqn\surs{\CF=\sum _{\CS} N^{\chi} \kappa_{0}^{\CN_{0}}\kappa_{1}^{\CN_{1}}}
where $\chi$ is the Euler characteristic of the surface, $\CN_{0}$ is
the number of its triangles mapped onto a single point, and $\CN_{1}$ is
the number of triangles mapped onto a link.

 Performing the gaussian integration  over the link variables
$A^{<xx'>}$ we find
\eqn\mtr{\eqalign{
\CZ=&\int \prod _{x} d\Phi^{(x)}e^{N\tr (- {1 \over 2}[\Phi^{(x)}]^{2}
+{\kappa_{0} \over 3}[ \Phi^{(x)}]^{3})} \cr
&\prod_{<xx'>}
{\det }^{-1}[I\otimes I -{ \kappa_{1} \over 2}
(I \otimes \Phi^{(x)} - \Phi^{(x')}
\otimes I)]\cr}}
where by $I$ we denoted the $N\times N$ unit matrix.

Because of the gauge symmetry $\Phi^{(x)} \to U_{x}\Phi^{(x)}U_{x}^{-1}$ the
partition function depends only on the radial degrees of freedom,
the eigenvalues % $z_{1}^{(x)}, ..., z_{N}^{(x)}$
of the matrix $\Phi^{(x)}$.
It is convenient to perform, together with the diagonalization, a linear change
of variables
\eqn\chng{
\Phi^{(x)}_{ij} = {1 \over \kappa _{1}}\delta _{ij}
 + a \ \sum _{k=1}^{N} U^{(x)}_{ik} U^{(x)}_{jk} z_{k}^{(x)}}
where
the small positive constant $a$ plays the role of a cutoff.
 Introducing the
potential
\eqn\vvp{ V(\phi)= -{N \over 2}({1\over \kappa_{1}}+az)^{2}
 + { N\kappa_{0} \over 3 } ({1\over \kappa_{1}}+az)^{3}}
we can write the matrix integral \mtr\ in the form
\eqn\prtp{\CZ=\int \prod_{x\in X} \prod _{j=1}^{N}dz_{j}^{(x)}
 \ \ e^{ - \CA
[z]}}
where
\eqn\acc{\eqalign{
 \CA
[z]=& -\sum _{x,j} V(z_{j}^{(x)}) +
\sum_{x,x'\in X}
 \ \sum_ {j,k =1}^{N} [C_{xx'} \log |z_{j } ^{(x)}+ z_{k} ^{(x')}|]\cr
 &-\delta _{x,x'} (1-\delta_{jk})
\log |z_{j}^{(x)}-z_{k} ^{(x')}| \cr}}
This integral is divergent but can be given meaning as a
formal series in $1/N$.
The action \acc\ depends on the two parameters $\kappa_{0}$ and
$\kappa_{1}$ as well as on the cutoff $a$ through the potential $V(z)$.

The $1/N$ expansion can be performed either as a quasiclassical expansion
for the collective field - the eigenvalue density of the  matrix  $\Phi$.
%or by reformulating the problem as a system of free fermions in a common
%potential  \ref\bipz{E. Br\'ezin, C. Itzykson, G. Parisi and J.-B. Zuber,
%{\it Comm. Math. Phys.} 59 (1978) 35}.
We will explain the method
 avoiding some
technical points arising in passing to the continuum limit
  which can be reconstructed from the text of ref. \Idis .

The effective dimension $C$ of the target space (= the central charge of the
matter fields) is related to the largest eigenvalue $\beta$ of the connectivity
matrix \conm\ by
\eqn\poiu{ C=1-6{(g-1)^{2} \over g}, \ \ \ \beta = 2 \cos (\pi g), 0<g<2  }
The continuum limit is achieved along a line in the $(\kappa_{0},
 \kappa_{1})$-space where one has to choose the branch $0<g< 1$.
At the endpoint of this line the critical singularity changes
and  is described by the
branch $1 \le g \le 2$ \Idis .

The double scaling limit is achieved if $N$ goes to infinity
 along the trajectory
\eqn\dsl{a^{1+g} N=\kappa}
where $\kappa$ is the renormalized string interaction constant.
In this limit the model can be studied by means of the collective field
method of Das and Jevicki \ref\djev{S. Das and A. Jevicki, {\it Mod. Phys.
Lett. }A5, 1639 (1990)}.

\newsec{The collective field  method}

In the large $N$ limit it is convenient
 to replace the integration in the space of
the eigenvalues $z_{j} ^{(x)}$ by a
 functional integration with respect to the
spectral density
\eqn\spd{\rho_{x} (z)=
 \sum_{j=1}^{N} \delta ( z - z_{j}^{(x)})
, \ \ \ x \in X}
The Jacobian for the change of variables
\eqn\chv{\int \prod _{j=1}^{N}
dz_{j}^{(x)}=\int \CD \rho ^{(x)} J^{(x)}[\rho^{(x)}]}
can be represented as a functional  integral over a  Lagrange multiplier
 field $\alpha_{x}(z)$, according to the suggestion of G. Parisi developed
by Migdal in
 \ref\miqcd{A. Migdal, 1/N  expansion and particle
spectrum in induced QCD, Princeton preprint PUPT-1332, July 1992}
\eqn\jacc{\eqalign{
&J^{(x)}[\rho_{x}]= \cr
&\int \CD \alpha _{x} \prod _{j=1}^{N}
 dz_{j}^{(x)}  \exp  \Big(  \sum_{x\in X} \int dz \
\alpha_{x} (z) [-\rho_{x} (z) + \sum _{j=1}^{N}
\delta ( z - z _{j}^{(x)} )] \Big) \cr
&= \int \CD \alpha_{x} \exp \Big(  \sum_{x\in X}\big[- \int dz
\ \alpha _{x} (z) \rho_{x} (z)
+ N \log \big(\int dz\  e^{ \alpha_{x} (z)}\big) \big]\Big) \cr}}
Now the partition function \acc\ is given by the  functional
integral
\eqn\ffi{ \CZ = \int \CD \rho \CD \alpha \  e^{-\CA _{eff}[\rho , \alpha]}}
with
\eqn\aaca{\eqalign{
&\CA_{eff} [\rho , \alpha]
 =\sum_{x \in X}\int dz \rho_{x} (z) [-  V(z)
+ \alpha_{x}(z )]
- N\sum_{x\in X} \log[\int dz e^{\alpha_{x}(z)}]\cr
&+ \sum_{x,x' \in X} \int dz
 \rho_{x} (z) \int dz '\rho_{x'} (z ')
[C_{xx'} \log|z + z '| - \delta _{x,x'} \log |z - z '|]\cr}}
The $1/N$ expansion  is the  quasiclassical
expansion for the field theory with action \aaca .  Let us denote by
 $\bar \rho_{x} (z), \bar \alpha_{x}(z)$
the classical solution minimizing this action.We
parametrize the quantum fluctuations around $\bar \rho_{x} (z)$ and $\bar
\alpha_{x}(z)$ as
\eqn\pps{\rho_{x} (z)= \bar \rho_{x} (z)
 -    {d \over dz}\phi _{x}(z), \ \ \
 \alpha_{x} (z)=\bar \alpha_{x} (z) +i \e_{x}(z)}
where the  function $\phi(z)$  vanishes
for $z \to \pm \infty$. Then the  the normalization condition
\eqn\noco{ \int _{-\infty}^{\infty} dz \rho_{x} (z)= N}
is automatically satisfied. We are then obliged to  introduce
a gauge condition supressing the translational zero mode of $\alpha $
\eqn\gcc{
 \int dz \bar \rho _{x}(z) \e _{x} (z) = 0}

Taking the  first derivative of the action with respect  to
the fluctuations $\phi$ and $\e$
we find the classical equations
\eqn\speq{{dV(z) \over dz} +  \sum _{x'\in X}
 \int dz '\bar \rho_{x} (z ')
[2\delta _{x,x'} {1 \over z - z '} -C_{xx'} {1 \over z + z '}]=0}
and
\eqn\epps{\bar \alpha_{x} (z) =  \log  \bar \rho_{x} (z) +{\rm constant}
}
The constant is fixed to be zero by the gauge condition \gcc . The advantage
of this choice is that
in the large $N$ limit the logarithm in \jacc\ can be replaced by the first
term in its expansion
\eqn\llof{N \log\Bigg[1- {1 \over N} \int dz \bar \rho_{x} (z)\Big[1-
e^{i\e_{x} (z)} \Big] \Bigg]
\to  - \int dz \bar \rho_{x} (z)\Big[ 1-
e^{i\e_{x} (z)}  \Big]
}

The new effective action reads
\eqn\efac{\eqalign{
\CA_{0}[\phi , \e , \xi]=& \sum _{x,x'\in X} \int dz
 \int dz ' \phi_{x} (z) \Big[{C_{xx'} \over (z + z ')^{2}}
+{\delta_{x,x'} \over (z - z ')^{2}}\Big]\phi _{x'}(z ')\cr
& + \sum _{x\in X} \int  dz
 [\log \bar \rho _{x} (z) +i \e_{x}(z)]
{d\phi_{x}(z) \over dz} + i \sum _{x\in X} \xi_{x}
 \int dz \bar \rho (z) \e (z) \cr
& -  \int dz \bar \rho_{x} (z)
e^{i\e_{x} (z)}
\cr}}
where $\xi_{x}$ is a Lagrange multiplier for the gauge condition \gcc.
In the continuum limit $a \to 0$ the theory depends on the  renormalized
cosmological
constant $\m$ and the string coupling $\kappa$ through the dimensionless
combination $\kappa ^{2} \m ^{1+g}$; in our normalization $\kappa =1$.
The dependence on $\m$ is
through the classical spectral density $\bar \rho _{x}(z) $ which will be
calculated below.

\newsec{ Classical solution}

We will confine ourselves to the case of target space $X=\Z$
of the SOS string where
\eqn\ccmc{C_{x x'}= \delta _{x, x'+1}+\delta _{x, x'-1}}
In certain sense this target space contains all other one-dimensional target
spaces.

We will solve eq. \speq\ assuming that the spectral density is supported
by the interval $-\infty <z<-\mm$. Then $\bar \rho _{x}(z)$ can be obtained
as the discontinuity of the analytic function
\eqn\wwwc{ \bar w_{x}(s)=
 {1 \over \pi}\int _{- \infty}^{-\mm} dz' {\bar \rho_{x}(z')
\over z - z '} }
 along the cut.
\eqn\axac{\bar \rho _{x}(z)= {1 \over \pi} {\rm Im}\  \bar w_{x}(z)}
The integral equation \speq\ then implies
\eqn\ocod{\eqalign{
{d V(z) \over z} + 2{ {\rm Re}} \ \bar  w_{x}(z)
+ \sum_{x'}C_{xx'} \bar w_{x'}(-z)& =0, \ \ \  -\infty < z < -\mm \cr
{{\rm Im}} \ \bar w_{x}(z) &=0, \ \ \  z > -\mm \cr  }}
The potential term $dV/dz$ containes no singularities in the
limit $a\to 0$
 and therefore
 can be
elliminated by a shift of the $\bar w$.  It only determines the position of
the cut.

It is convenient to transform the cut $z$-plane into a semi-infinite strip
$\{ \r \tau \ge 0,| \i \tau | \le \pi \}$ by a  change of
variables
\ref\Icar{I. Kostov, ``Strings embedded in Dinkin diagrams, in the  Proc.
of the Cargese meeting {\it Random Surfaces, Quantum Gravity and Strings},
Saclay Preprint SPhT/90-133}
\eqn\prms{z = \mm \cosh \tau . }
Then the two sides of the cut are parametrized by the boundaries
$\{ \tau \pm i\pi , \tau >0 \} $ of the strip.
Now  eq.\ocod\ can be written in the form
\eqn\looe{\bar w_{x}(\tau + i \pi) +\bar w_{x}(\tau - i \pi)
+ \bar w_{x+1}(\tau) + \bar w_{x-1}(\tau)=0}
The general solution of this equation is a superposition of plane waves
\eqn\sogu{ \bar w_{(\nu)}(\tau) = M^{\nu} \cosh
  ( \nu \tau ) e^{ipx};
\ \   -1<p<1, \ \nu = p + 2m+1,  m \in \Z}
One can show \Idis\ \ref\Imat{I. Kostov and M. Staudacher, Saclay  and Rutgers
Preprint SPhT/92/025, RU-92-6, to appear in \np }
that the physical solution is
\eqn\plop{\bar w (\tau) = - {d \over dp} \bar w_{(1+p)}(\tau) |_{p=0} =
- \mm \tau \cosh \tau + {\rm regular \  terms}}
\eqn\rrrh{\bar \rho (\tau) = \mm \sinh \tau = \sqrt{z^{2}-\m}}

In the case of a general target space $X$
one should diagonalize the connectivity matrix
\eqn\diia{C_{xx'}=\sum _{p \in P}S_{(p)}^{x} \ 2\cos (\pi p)\  S_{(p)}^{x'}}
where  $P$  is the momentum space dual to $X$.
Then the classical solution is given by the plane wave with the minimal
momentum $p_{0}=|g-1|$
\eqn\poi{ \bar \rho _{x}(z) \sim S_{(p_{0})}^{x}(\mm)^{g}
 \sinh (g\tau)}
All other  other plane wave solutions  are down by
positive powers of the cutoff $a$.

\newsec{The kinetic term}

The quadratic kernel in the action \efac \ is essentially the inverse
propagator in the string field theory.
It is given by
\eqn\weqn{\eqalign{
& [\hat G^{-1} \phi]_{x}(z) ={\delta \CA_{0} \over \delta \phi_{x}(z)}\cr
&= -{\p \over \p z}\sum_{x'}
\int _{-\infty}^{-\mm}dz' \Big[ {2\delta_{xx'} \over z-z'}
+{C_{xx'} \over z+z'}\Big]\phi_{x'}(z') +\log \bar \rho _{x}(z)
 \cr}}
 If we  are interested only in the perturbative expansion
 in $\kappa$, we can assume that the quantum field $\phi$ also
vanishes for $z>- \mm$. In this case we can
  introduce   the analytic fields
\eqn\diisr{\psi (x,\tau) = {1 \over \pi }
\int _{-\infty}^{-\mm}  {dz'
\over z - z '}\phi _{x}(z ')  ,\ \  \ \ z =\mm \cosh \tau }
related to the field $\vp (x, \tau) \equiv \phi_{x}(-\mm \cosh \tau)$ by
\eqn\varra{ \vp (x,\tau ) = \sin (\p _{\tau})
\psi (x, \tau)}
It will be convenient to consider the coordinate $x\in \Z$ as a continuum
parameter.
The  inverse propagator \weqn\  then is represented by the
differential operator
\eqn\qdrp{ \hat G^{-1} =
(2\cos (\pi  \p_{\tau}) - 2\cosh \p_{ x})\
{\p_{\tau}\over  \sin (\pi  \p _{\tau})}}
which is diagonalized by plane waves
\eqn\plw{\langle E,p|\tau ,x\rangle = \cos (E\tau)e^{i\pi px}; \ E\ge 0,
-1\le p <1}

In the case of a general target space the kinetic term reads
\eqn\diah{\langle \phi|\hat G^{-1}|\phi\rangle =
 \sum _{p\in P}\int _{0}^{\infty} dE \  \vp(E,p) {E
 \over \sinh (\pi E)}  [\cosh (\pi E)
-\cos (\pi p)] \vp(E,p)}
This quadratic form produces exactly the string propagator which we
have found in \Idis\ (eq. (4.52)) by means of the loop diagram technique.

Knowing the kinetic term one can easily calculate the
correlation function of the loop operators
\eqn\twlp{w_{x}(\l)= \int _{-\infty}^{-\mm} dze^{\l z} \phi_{x} (z)
=\int _{0}^{\infty} d\tau  e^{-\mm \l \cosh \tau}{\p \over \p \tau}
\vp(x, \tau)}
It can be represented as an integral over Bessel functions
\eqn\loopi{\eqalign{
&\langle w_{x}(\l)w_{x'}(\l')\rangle  =\cr
&=\sum _{p \in P} \int  _{0}^{\infty} dE\  S_{(p)}^{x}
K_{iE}(\mm \l ){E\sinh (\pi E)
\over \sinh (\pi E) - \cos (\pi p)}  S_{(p)}^{x'} K_{iE}(\mm \l ')
\cr}}

The two-loop correlator in $C=0, C=1/2$ and $C=1$
strings has been considered from the viewpoint of
 an effective string field theory in refs.
\ref\mss{G. Moore, N. Seiberg and M. Staudacher, \np 362 (1991) 665},
\ref\mooo{G. Moore, \np 368 (1992) 557}, \ref\ms{G. Moore and N. Seiberg,
{\it Int. Jour. Mod. Phys.} A7 (1992) 2601}.
The expression obtained in the momentum space using the formalism of
the large-$N$ matrix quantum
mechanics was different than the integrand in  \loopi .
 Now this contradiction is resolved:
the discrepancy comes from the fact that the two correlators are
considered different momentum spaces, a compact and noncompact one
 \ref\mimi{I. Kostov and M. Staudacher,
Saclay and Rutgers Preprint SPhT/92-092, RU-92-21, submitted to \pl }.
One can easily check that in the  $x$-space the loop-loop correlators
 coincide, using the identity
\eqn\edhhe{\eqalign{
 &\langle x_{1}| {1 \over
2\cosh ( \pi E) - 2\cos \p_{ x}} \
{E \over  \sinh (\pi E) } |x_{2}\rangle \cr
&=  \langle x_{1}| {1
\over \pi ^{2} E^{2}+ \p_{x}^{2}}|x_{2}\rangle \cr
& = {1 \over E}
e^{-\pi E|x_{1}-x_{2}|}
, \  \  \ \ \ x_{1}-x_{2}\in \Z \cr}}
Amusingly, our string propagator restricted
 to the $x$-space is just the propagator of a
two-dimensional Euclidean massles particle.
\newsec{Discussion}

The matrix model considered here gives a unified
description of all string theories with rational central charge
$C$ of the matter fields.
The collective field method should work equally well for $C<1$ and $C=1$.
Up to two loops it yields the same results
 as those obtained by the loop gas method \Iade , \Idis .

At the moment we do not know an effective method for
performing the integration over the Lagrange multiplier field $\e $.
It is however evident that this would  lead to a {\it local} in the
target space interaction.
This means that
 topology changing processes thus occur at a single point of the
target space.  The effective potential
 reproducing the loop diagram technique of ref.\Idis\
(Eqs. (4.51) - (4.53))
should have  the form
\eqn\eehmatushkarus{\CV[\vp ] = \sum _{x\in X}\CV_{x}
 (\p _{\tau}\vp(x, \tau)|_{\tau =0},
\p _{\tau}^{3}\vp(x, \tau)|_{\tau =0}, ... )}
We intend to consider the string interactions in a future publication.

Finally, let us comment briefly the relation between our model and the
matrix quantum mechanics which is the most studied model of the $C=1$ string.
If we return to the original definition \partf - \actn\
of the matrix model and perform the integration over the
$\Phi$-matrices  (take the simplest case  $\kappa_{0} =0$)
we will obtain a particular case of the matrix chains considered in ref.
\Dal . It is defined by the transfer matrix
\eqn\trm{K(y,y')= \exp [(y+y')^{2} ]}
which can be thought of  as a singular limit of the inverse oscillator
kernel at time interval which is half of the self-dual radius.
This explains why the spectrum of the propagator \qdrp\ does not contain
poles at integer momenta corresponding to the so-called  ``special states''.
 The half-wave-lengths of the
 ``special states'' are multiples of the
 distance between two subsequent points of the discrete target
space $\Z$. Therefore these states are inobservable.

\bigbreak\bigskip\bigskip\centerline{{\bf Acknowledgements}}\nobreak

The author thanks Mike Douglas, V. Kazakov, A. Migdal and M. Staudacher
for discussions, and V. Pasquier for a critical reading of the manuscript.

\listrefs
\bye
***********************************************************
\newsec{ Reduction to a problem of  free fermions}
Let us return to the original representation \partf -\actn\ of the
partition function and decompose each $A$-matrix as a product of a diagonal
matrix

with non-negative elements and two $U(N)$ rotations
\eqn\decc{\eqalign{
&A^{<xx'>}= {1 \over a N \kappa_{1}}
 U^{<xx'>}\sqrt{\l^{<xx'>}}V^{<xx'>} ,\cr
&\l^{<xx'>} =\l^{<x'x>}_{r} = diag (\l^{<xx'>}_{1},...,\l^{<xx'>}_{n})\cr}}
The integration w. r. to the unitary matrices can be readily performed
using the Itzykson-Zuber formula \ref\IZ{C. Itzykson and J.-B. Zuber,
{\it J. Math. Phys. } 21 (1980) 411} and the result is
\eqn\npfc{\eqalign{
\CZ =& \prod _{x\in X}\prod _{j=1}^{N} \int dz_{j}^{(x)}
  e^{N V[z_{j}^{(x)}]} \Big(\Delta [z^{(x)}]\Big)^{2}\cr
&\prod_{<xx'>}\prod_{j=1}^{N}
\int _{0}^{\infty} d\l^{<xx'>}_{j}
{{\det}_{jk} \Big[ e^{ \l^{<xx'>}_{j} z_{k}^{(x)}}\Big]
{\det}_{jk} \Big[ e^{\l^{<xx'>}_{j}z_{k}^{(x')}}\Big]
\over \Delta [z^{(x)}]\Delta [z^{(x')}]}\cr}}
where we have used the notations \chng\ and \vvp\ , and
\eqn\vabb{\Delta (z)= \prod_{j<k} (z_{j} - z_{k})}
is the usual Van-der-Monde deterninant and the product goes over the
nonoriented links $<xx'>$.
We see that the Van-der-Monde deterninants cancel along each linear
branch of $X$.
Let us concentrate on the case of the SOS string when $X=\Z_{n}$.
Then there are no Van-der-Monde deterninants at all.
If we denote $\l^{<x>}_{j}=\l^{<x-1, x>}_{j}; j=1,...,N, \ x=1,...,n$,
the partition function is given by
\eqn\ttrr{\eqalign{
\CZ _{n} =&
\int\prod_{j=1}^{N} dz_{j}^{(0)}dz_{j}^{(1)}...dz_{j}^{(n)}
{\det} _{jk}\Big[ \delta (z_{j}^{(0)} - z_{k}^{(n)}) \Big]
e^{N(V[z _{j}^{(1)}]+...+V[z _{j}^{(n)}])}\cr
& \int_{0}^{\infty} d\l^{<1>}_{j}...d\l^{<n>}_{j}
 \ \exp \{[z_{j}^{(0)}+z_{j}^{(1)}]\l^{<1>}_{j}+
[z^{(1)}_{j}+z^{(2)}_{j}]\l^{<2>}_{j}+
 ... \cr
& +[z^{(n-1)}_{j}+z^{(n)}_{j}]\l^{<n>}_{j}\}\cr}}
  The antisymmetrization
can be done at one point $x$ only and any other determinant
 can be replaced by the first ordering in its expansion.
The integration over the $\l$-variables can be performed immediately
and the result is
\eqn\pprt{\eqalign{
\CZ_{n}=& \prod_{j=1}^{N} \int d\phi_{j}^{(0)}...d\phi_{j}^{(n)}
{e^{NV[\phi^{(1)}_{j}]+...+NV[\phi^{(n)}_{j}]} \over
(\phi ^{(0)}_{j } + \phi^{(1)} _{ j })(\phi ^{(1)} _{ j } + \phi^{(2)} _{ j })
...(\phi^{(n-1)} _{ j } + \phi^{(n)} _{ j })}\cr
&{\det}_{jk}\big[ \delta (\phi^{(0)}_{j}-\phi^{(n)}_{k})\big]\cr}}
We see that the Van-der-Monde determinants in the integrand of \prtp\
effectively disappear after antisymmetrization and what is left is a kind of
sinh-Gordon chain.

In order to find a connection with the standard matrix chain
\ref\paiir{G. Parisi, \pl \  238B (1990) 209}, we perform the integration
in the opposite order: first integrate over the $\phi$-variables, and then
over the $\l$-variables. The result of the first integration is
\eqn\mtrch{\eqalign{
\CZ_{n}=& \int_{0}^{\infty}\prod_{j=1}^{N}
d\l^{<1>}_{j}... d\l^{<n>}_{j}{\det}_{jk}\big[K(\l^{<1>}_{j},\l^{<2>}_{k})
\big]
...    \cr
&{\det}_{jk}\big[K(\l^{<n-1>}_{j},\l^{<n>}_{k})\big]
{\det }_{jk} \big[K(\l^{<n>}_{j},\l^{<1>}_{k})\big]
\cr}}
where
\eqn\krnl{K(\l , \l ')=\int d\phi
e^{NV(\phi)+\phi(\l + \l')} = e^{-f(\l + \l ')}}
Since we are interested in the double scaling limit$N \to
\infty , a \to 0 ; a^{2}N=\kappa$, only the gaussian part of the potential is
important. Writing explicitely the cubic potential \vvp\
\eqn\cbc{NV(\phi)= -{1 \over a} {\kappa_{1}-\kappa_{0} \over
\kappa_{1}^{2}} -\kappa ({1 \over 2}- {\kappa_{0} \over \kappa_{1}})
\phi^{2} +{1 \over 3} \kappa \kappa_{1} a \phi^{3}}
 we see that the cubic term is irrelevant. Thus , after a linear change
of variables
\eqn\chng{\eqalign{
& \l = {c \over a} + A\lambda ,\cr
 &c =   \kappa
 (\kappa - \kappa_{0})/\kappa_{1}^{2},
\ \ A=\sqrt{\kappa (1-2\kappa_{0}/ \kappa_{1})} \cr} }
the kernel \krnl\ can be simplified to
\eqn\skr{K(\lambda, \lambda ') = e^{{1 \over 2} (\lambda + \lambda ' )^{2}}}
but the integration is restricted to the semiinfinite interval $  \lambda >
-c/a$.
The constant $c$ can be elliminated by a redifinition of $a$ and will be
assumed to be 1. Again all the
 determinants but one can be replaced by the term with
trivial ordering and finally one obtains  in the scaling limit
\eqn\sclm{\eqalign{
\CZ _{n} =&
\int_{-1/a}^{\infty} \prod_{j=1}^{N} d\lambda_{j}^{(0)}d\lambda_{j}^{(1)}
...d\lambda_{j}^{(n)}
{\det} _{jk}\Big[ \delta (\lambda_{j}^{(0)} - \lambda_{k}^{(n)}) \Big]
\cr
& \exp \{{1 \over 2}[(\lambda_{j}^{(0)}+\lambda_{j}^{(1)})^{2}+
(\lambda^{(1)}_{j}+\lambda^{(2)}_{j})^{2}+
 ... +(\lambda^{(n-1)}_{j}+\lambda^{(n)}_{j})^{2}]\}\cr}}
This is the partition function of the ``Weingarten model \`a
la Polyakov''  \Dal\ for a special choice of the parameters.
The model considered in \Dal\ is in turn equivalent to the
ordinary (non gauge-invariant) matrix chain
 for the random surface embedded in a discretized
circle \ref\grkl{D. Gross and I. Klebanov, \np 344 (1990) 475 and
B354 (1991) 459 } \ref\bkz{D. Boulatov and V. Kazakov,
``One-dimensional string theory with vortices as the upside-down
matrix oscillator'', preprint LPTENS 91/24, KUNS 1094
HE(TH) 91/14, August 1991} if the effective distance between two
subsequent points is less than the critical one. As we have conjectured in
\Idis , the partition function \sclm\ corresponds exactly to the critical
distance where a Kosterlitz-Thouless transition is expected to
 occur due to the
liberation of the vortex configurations related to the angular degrees
of freedom.

In the gauge invariant model there
are no vortices and therefore no transition. Nevertheless there is a
scale in the target space which is defined by the minimal wavelength
of the so called discrete states; it is equal to the half of the
critical distance.
 The  $\Z_{n}$ string corresponds to
the continuum string with radius of compactification $R=n/2$. Note that
$n$ should be even if we want to have a random surface interpretation
of the matrix model.

It is known that the ordinary $D=1$ string is equivalent to a system of
noninteractting fermions moving in an inverse oscillator potential.
As was mentioned in \ref\kleb{I. Klebanov, String theory in two dimensions,
PUPT-1271, July 1991} and \Dal , the discrete matrix chain with general
potential can
 be mapped onto the same problem, if the lattice spacing does not exceed
some critical value. Let us see whether this correspondence worls in our
model.

The general gaussian kernel $K(y_{1} , y_{2})$ can be represented as
the evolution kernel of an oscillator potential
\eqn\osccc{\langle y_{1}|e^{{1 \over 2} T (\p _{y}^{2}+\omega ^{2} y^{2})}
|y_{2}\rangle = \sqrt{\omega \over 2 \pi \sin (\omega T)}
\exp \Bigg[{\omega \over 2} {2y_{1}y_{2} - \cos (\omega T)
[(y_{1})^{2}+(y_{2})^{2}] \over \sin (\omega T)}\Bigg]}
Our kernel \skr\ can be obtained from \osccc/ as the limit
\eqn\ags{\omega T = \pi - \e, \omega = \e ; \ \ \ \ \ \e \to 0}
%Therefore the partition function \sclm\ is identical to the partition
%function of $N$ {\it free} fermions ($\omega$ =0, no harmonic potential),
%living on the circle with radius $R=n/2$
\eqn\pptr{\CZ_{n}=\int \CD y \det \Big[\delta(y_{i}(0) - y_{i}(2\pi R))
\Big] \exp^{ -{1 \over 2} \int _{0}^{2\pi R} dx [ \big( {dy \over dx}\big)^{2}
-y^{2}] }}
In the case of general $R$ the partition function can be obtined by
analytic continuation from the ordinary oscillator \bkz\

\eqn\bka{\CZ ^{(N)}(R)= {e^{- {1 \over 2} i \pi R N^{2}}
\over (1-e^{-i\pi R})(1-e^{-i\pi 2 R})...(1-e^{-i\pi N R})}}
 This formula is singular when $R$ approaches integer values.
In order to avoid the artificial singularity we introduce, as usual,
 a chemical potential $\m$
\eqn\bkb{\CZ (\m , R)= \sum_{N=0}^{\infty} \CZ^{N}(R) e^{\m R N^{2} \pi}
=\exp \sum_{k=0}^{\infty} \log \{1+e^{\pi R[\m - i(k+1/2)]}\}}

As a function of $\m$ the partition function behaves smoothly in $R$
\eqn\bkc{\eqalign{
\log \CZ (\m , R)
 =& {1 \over \pi} \int _{-\infty}^{\infty} dE
\sum_{k=0}^{\infty} {k+1/2 \over E^{2}+(k+1/2)^2} \log \{1+e^{\pi R(\m -E)}\}
\cr
= &\int_{-\infty}^{\infty} dE \rho (E) \log \{1+e^{\pi R(\m -E)}\}\cr}}

\eqn\bkd{\rho (E) = - {1 \over 2 \pi} \r \psi[iE+{1 \over 2}] + {1 \over \pi}
\log {1 \over a^{2}} ={1 \over 2\pi} \i \int _{0}^{\infty} d\tau
{e^{iE\tau } \over \sinh  \tau /2}}

The relation between the Fermi level $\m$ and the cosmological constant
$\kappa = Na^{2}$ is given by the normalization condition
\eqn\bke{{\p \over \p \m} \log \CZ (\m , R)
 = 2\pi  NR\int_{-\infty}^{\infty}dE \rho (E)
{1 \over 1+ e^{\pi R (\m - E)}}}

\eqn\bkf{{\p ^{3}  \over \p \m ^{3}} \log \CZ (\m , R)
= {R \over 4} \i \int _{0}^{\infty}
 d\tau
 {\tau /2 \over \sinh \tau /2} { \tau /R \over \sinh \tau /R} e^{i\m \tau }}
\eqn\bkg{ {R  \to 4 /R, \m \to {R \over 2 } \m }}

% Instead of \listfigs, if \fig not used in text, after \listrefs say
%
%\figures       % former version, still supported
%\fig{1}{This is the first figure caption.}
%\fig{2}{This is the second figure caption.}
%\fig{3}{Note that \\fig automatically types a colon and lines up the
%text properly.}
%
%\parindent=20pt

\listrefs
%\listfigs   %(if necessary)
\bye